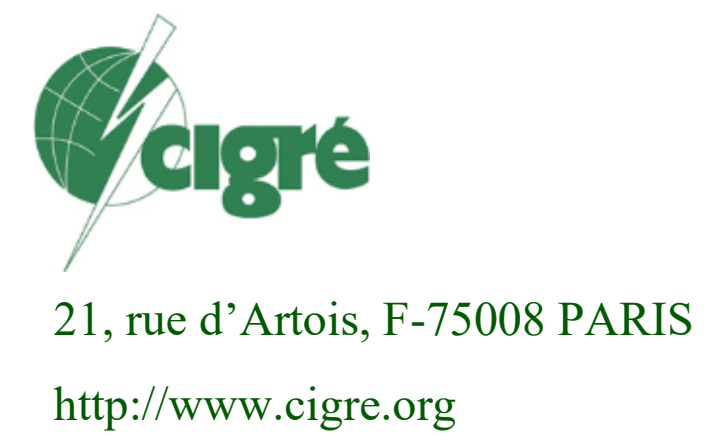

# Grid-Forming Enhanced STATCOMs for Wind Power Plant Stability: Design, Modeling, and Control

**P. R. BANA[†]*, J. P. HASLER[†], R. HEYDARI[†], A. J. OWENS[†], F. CHIUMINATTO[††]**
**Hitachi Energy**
**Sweden[†], USA[††]**

**SUMMARY**

The increasing integration of renewable energy sources (RES) into modern power systems, particularly in weak grid environments, demands advanced solutions for voltage and frequency stability. This paper demonstrates that Enhanced STATCOMs (E-STATCOMs) with grid-forming capabilities outperform synchronous condensers (Syncons) in wind farm applications. Using a power-admittance-based linear modeling framework and validated through detailed EMT simulations, the study shows that E-STATCOMs provide superior damping of sub-synchronous resonance (SSR), faster dynamic response, and effective suppression of DC components, transient overvoltages, and new oscillatory modes introduced by series compensation. Unlike Syncons, which are limited by mechanical inertia, E-STATCOMs offer tunable control, virtual impedance emulation, and scalable fault current support. These advantages make E-STATCOMs a more robust and flexible solution for stabilizing renewable-dominated grids, especially those with long transmission lines and series compensation. The paper also aims to provide a benchmark E-STATCOM design criterion that enables stakeholders, such as transmission system operators (TSOs), to assess its suitability without the need for exhaustive EMT simulations.

prabhat.ranjan-bana@hitachienergy.com

# 1 INTRODUCTION

The global transition toward renewable energy sources (RES) has significantly increased inverter-based resources (IBR), particularly wind and solar power [1-2]. This shift, while environmentally beneficial, introduces new challenges in maintaining grid stability, especially in weak grid environments with low short-circuit strength and reduced system inertia [3]. Traditionally, synchronous condensers (Syncons) have been deployed to provide partial inertia, voltage regulation, and fault current support. However, their mechanical nature limits their responsiveness and adaptability under dynamic grid conditions.

To address these limitations, transmission system operators (TSOs) are increasingly adopting Enhanced STATCOMs (E-STATCOMs) with grid-forming (GFM) capabilities [3-5]. These devices offer fast, tunable control and synthetic inertia, making them well-suited for mitigating sub-synchronous resonance and supporting voltage and frequency stability in renewable-rich systems. In this context, careful attention must be given to the design of GFM control of E-STATCOM, as its dynamic behavior plays a critical role in achieving these performance objectives.

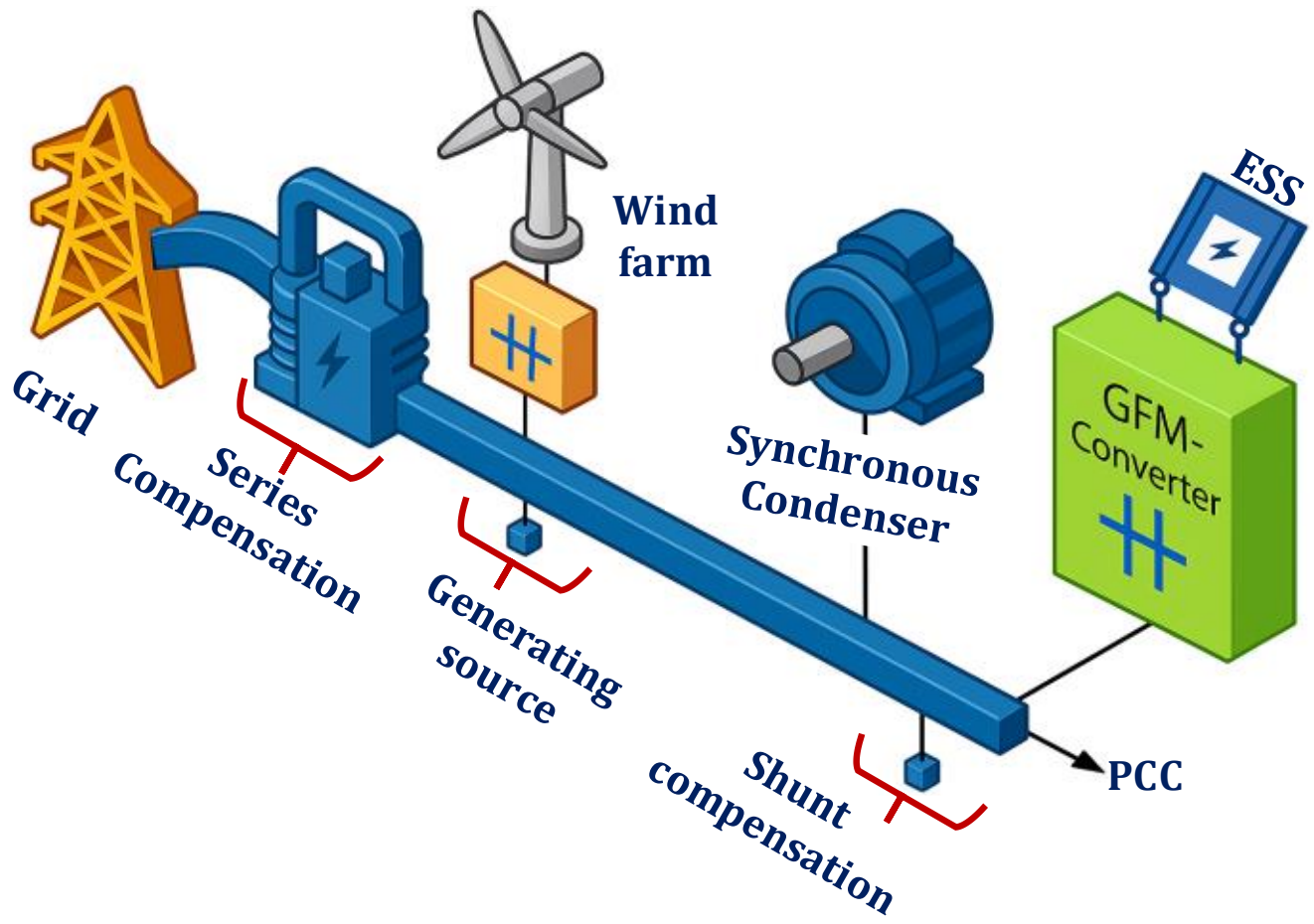

*Figure 1: Schematic of the overall system configuration*

The standard method for analyzing sub-synchronous resonance (SSR) and other small-signal issues in IBR-dominated systems, as shown in Figure 1, is the dq and sequence domain impedance analysis [6-7]. However, recent literature [7-8] has introduced power-response matrix-based modeling as a compelling alternative to conventional impedance-based approaches for analyzing converter dynamics. This method characterizes the converter system by directly relating variations in grid voltage magnitude and frequency to changes in active and reactive power output. Unlike dq-impedance models, which often require complex reference frame transformations, the power-response matrix formulation is inherently reference-frame independent, making it particularly suitable for aggregating multiple converter systems in large-scale networks.

Moreover, it enables a more intuitive assessment of GFM properties such as synthetic inertia, frequency droop, and damping behavior, key attributes for evaluating the performance of GFM-based E-STATCOMs in weak grid environments. This modeling approach aligns well with the objectives of this study, which seeks to develop a linear framework that captures the essential dynamic interactions between wind farms and grid-supporting devices.

In line with the above discussion, this paper utilizes a power admittance-based linear modeling framework to analyze and compare the small-signal dynamic response of Syncons and E-STATCOMs in wind farm applications. The study quantifies active and reactive power responses under various grid conditions by leveraging frequency-domain transfer functions. The goal is to capture the underlying complexities of grid-tied wind farm behavior. The resulting insights are intended to inform the design of E-STATCOMs and enhance system-level stability in high-renewable scenarios.

# 1 SYSTEM MODELLING AND ANALYSIS

The system configuration shown in Figure 1 is considered to study key challenges in renewable-dominated power networks and how to address them. It features a wind farm connected to the grid, with series compensation placed immediately after the grid to enhance power flow and voltage stability. Although the series compensation in the system can enhance certain services, it may also lead to other resonance issues, which will be detailed through the linear modeling approach. The shunt compensation units shall provide the damping service for this kind of resonance issue. In line with this, the setup further includes two alternatives for shunt compensation: a Syncon and an E-STATCOM, implemented via a GFM-converter. The aim is to evaluate the grid-supporting ability of these technologies, particularly in mitigating SSR, DC oscillations, and other grid-supporting services, and to design compensating devices accordingly.

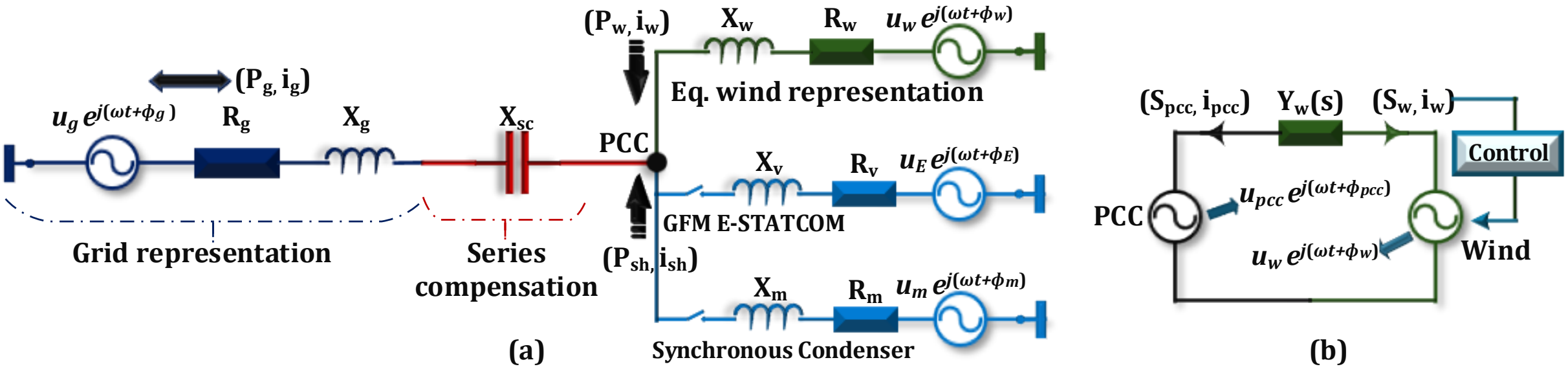

*Figure 2: Block diagram of (a) Grid-connected wind farm system with series and shunt compensation, (b) Equivalent circuit of wind connected to the PCC/infinite bus for linear modeling*

## 1.1 Grid-connected wind-farm system

The Thevenin equivalent circuit of the setup given in Figure 1 can be represented as in Figure 2(a). The subsequent linear modeling analysis is conducted based on this Thevenin equivalent system. For the purpose of streamlining the analysis, a single energy source (here, wind) connected to the infinite bus or PCC is considered, as illustrated in Figure 2(b). The wind farm is modeled as a voltage source ($\boldsymbol{u}_w$) connected through a physical reactor with admittance $\mathbf{Y}_w$ (comprising resistance $R_w$ and inductance $L_w$) to an ideal voltage source ($\boldsymbol{u}_{pcc}$), representing a strong grid with zero impedance.

In such a setup, the apparent power at the PCC, $\boldsymbol{S}_{pcc}$, is defined as follows:

$$\boldsymbol{S}_{pcc} = \boldsymbol{U}_{pcc}\boldsymbol{I}^*_{pcc} = \boldsymbol{U}_{pcc}\boldsymbol{Y}^*_w\left(\boldsymbol{U}_{pcc} - \boldsymbol{U}_w\right)^* = \boldsymbol{Y}^*_w u_{pcc} e^{j(\omega t + \varphi_{pcc})}\left(u_{pcc} e^{j(\omega t + \varphi_{pcc})} - u_w e^{j(\omega t + \varphi_w)}\right)^* \tag{1}$$

The above expression can be linearized by assuming the nominal voltage ($\boldsymbol{u}_{pcc} = \boldsymbol{u}_w = 1$p.u.) and no-load operating ($\phi_{pcc} = \phi_w = 0$) conditions, as follows:

$$
\begin{bmatrix} dP \\ dQ \end{bmatrix} = \underbrace{\begin{bmatrix} \frac{1/L_w\left(s+R_w/L_w\right)}{s^2+(2R_w/L_w)s+(R_w/L_w)^2+\omega_{s0}^2} & \frac{\omega_{s0}/L_w}{s^2+(2R_w/L_w)s+(R_w/L_w)^2+\omega_{s0}^2} \\ \frac{-\omega_{s0}/L_w}{s^2+(2R_w/L_w)s+(R_w/L_w)^2+\omega_{s0}^2} & \frac{1/L_w\left(s+R_w/L_w\right)}{s^2+(2R_w/L_w)s+(R_w/L_w)^2+\omega_{s0}^2} \end{bmatrix}}_{\begin{bmatrix} \mathrm{Re}(Y) & -\mathrm{Im}(Y) \\ \mathrm{Im}(Y) & \mathrm{Re}(Y) \end{bmatrix};\ \omega_{s0}=\omega \text{ is the nominal source frequency}} \begin{bmatrix} \left(du_{pcc}-du_w\right) \\ \left(d\varphi_{pcc}-d\varphi_w\right) \end{bmatrix} \tag{2}
$$

The admittance matrix in the above expression can be used for frequency-domain analysis akin to the method demonstrated in [7]. However, variations in initial conditions can shift the resonance frequency, reducing the accuracy of linear analysis. The following section will validate this statement through both frequency and time-domain results. Nevertheless, to address this, it is crucial to incorporate non-ideal initial conditions ($\boldsymbol{u}_{pcc} \neq \boldsymbol{u}_W \neq 1$p.u. and $\phi_{pcc} \neq \phi_W \neq 0$) into expression (1). Accordingly, the linearised admittance matrix can be reformulated as

$$
\begin{bmatrix} dP \\ dQ \end{bmatrix} = \begin{bmatrix} -y_r 2u_{pcc,0} + y_r u_{w,0}\cos(\Delta\varphi) + y_i u_{w,0}\sin(\Delta\varphi) & y_i u_{pcc,0}u_{w,0}\cos(\Delta\varphi) - y_r u_{pcc,0}u_{w,0}\sin(\Delta\varphi) \\ y_i 2u_{pcc,0} - y_i u_{w,0}\cos(\Delta\varphi) + y_r u_{w,0}\sin(\Delta\varphi) & y_r u_{pcc,0}u_{w,0}\cos(\Delta\varphi) + y_i u_{pcc,0}u_{w,0}\sin(\Delta\varphi) \end{bmatrix} \begin{bmatrix} du_{pcc} \\ d\varphi_{pcc} \end{bmatrix} +
$$
$$
\begin{bmatrix} y_r u_{pcc,0}\cos(\Delta\varphi) + y_i u_{pcc,0}\sin(\Delta\varphi) & -y_i u_{pcc,0}u_{w,0}\cos(\Delta\varphi) + y_r u_{pcc,0}u_{w,0}\sin(\Delta\varphi) \\ -y_i u_{pcc,0}\cos(\Delta\varphi) + y_r u_{pcc,0}\sin(\Delta\varphi) & -y_r u_{pcc,0}u_{10}\cos(\Delta\varphi) - y_i u_{pcc,0}u_{w,0}\sin(\Delta\varphi) \end{bmatrix} \begin{bmatrix} du_w \\ d\varphi_w \end{bmatrix} \tag{3}
$$

where $y_r$ and $y_i$ are the real and imaginary parts of $Y$ in (1); $u_{pcc,0}$, and $u_{w,0}$ are the initial values of $u_{pcc}$ and $u_W$, respectively.

## 1.2 Frequency and time-domain analysis

This section examines the power response of a voltage source with a physical reactor, reflecting a wind farm's expected subtransient behavior after a grid or network disturbance [9-10]. The expression in (2) is considered to model the wind farm and connect it to the grid. Figure 3(a) illustrates the frequency response of the wind power ($P_W$) to its corresponding phase angle perturbations for various values of the reactor resistance ($R_W$). The response exhibits the characteristics of a second-order low-pass filter, with two resonance frequencies; one around 10 Hz, referred to as the SSR and usually introduced by the power controller of the source. The other resonance occurs at the frequency aligned with the fundamental grid frequency of 50 Hz. This can also be referred to as the DC-component in the dq-frame and is usually introduced due to the phase reactor. The damping of this 50Hz component can be provided through the resistance or loss of the system.

Further, Figure 3(b) shows the time-domain results of $P_w$ to a corresponding $10^0$ phase angle jump. The power response of the wind-farm during a phase step event is inversely proportional to the phase reactor, $X_W$, showing a rise time of 1–2 cycles and weak DC-component damping when resistance is low.

As mentioned earlier, the system's initial conditions influence the resonance frequencies. To validate this, expression (3) is considered for frequency and time-domain analysis. From Figure 4(a), it can be observed that with the increase in initial power, the SSR frequency decreases, and the DC-component dampens. This is also evident from the time-domain waveforms presented in Figure 4(b), where the magnitude and impact of the DC-component are relatively less with higher initial power. Further, the SSR is shifted to a relatively lower frequency value.

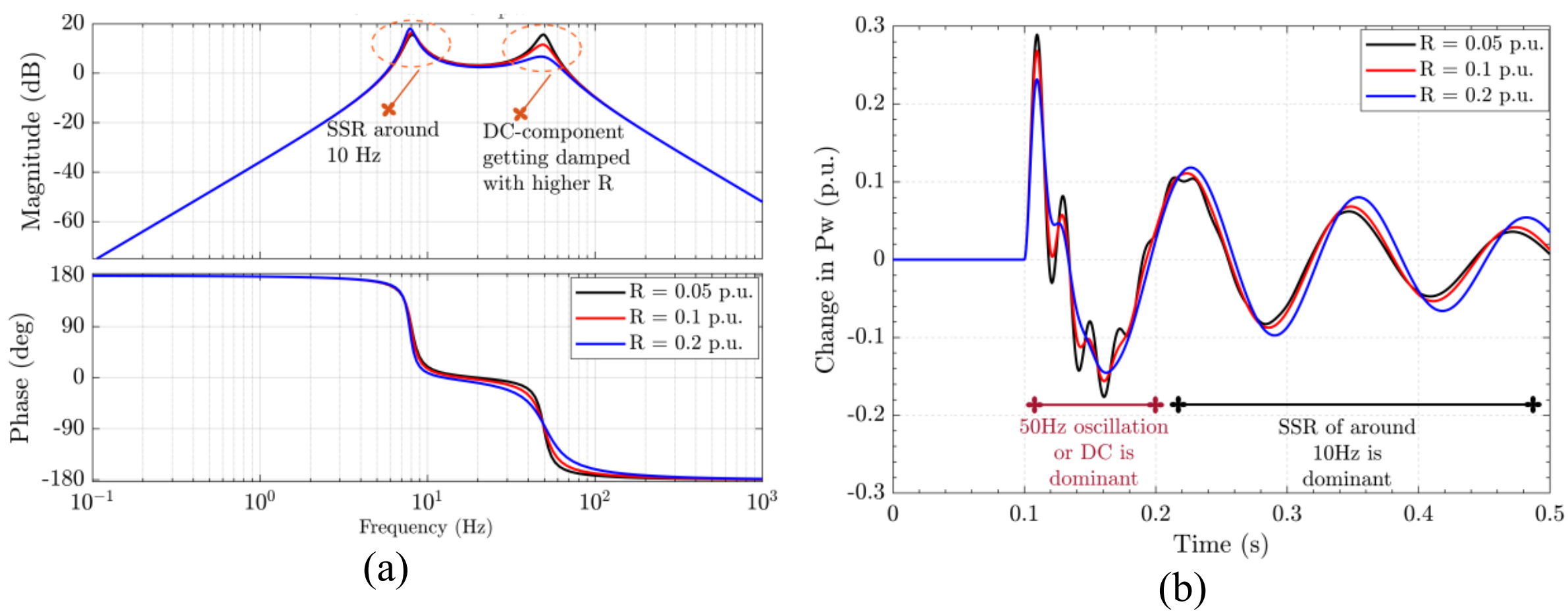

*Figure 3: Power response of the grid-connected wind farm system using expression (2) for $X_w$ = 0.5 p.u. and different $R_w$, (a) Frequency-domain results of $P_w$ with phase perturbations, (b) Time-domain response of $P_w$ with $10^0$ angle jump.*

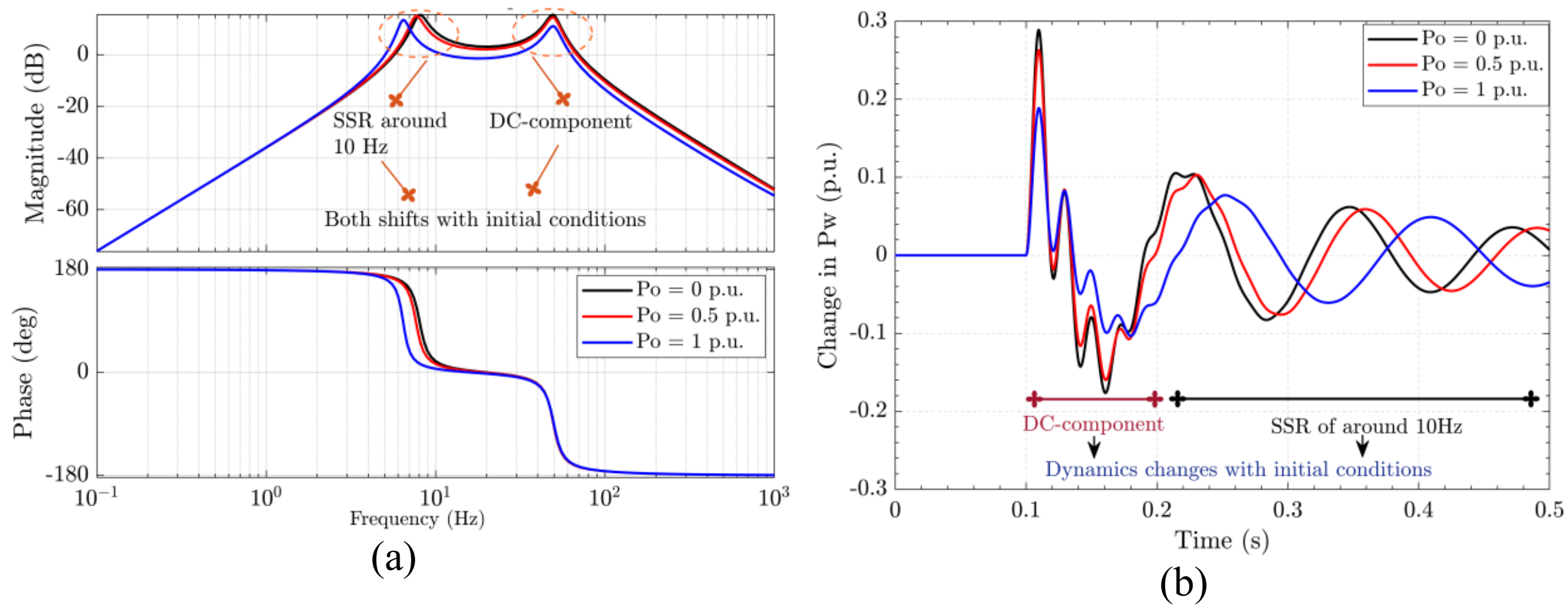

*Figure 4: Power response of the grid-connected wind farm system using expression (3) for different initial operating conditions, (a) Frequency-domain results of $P_w$ with phase perturbations, (b) Time-domain response of $P_w$ with $10^0$ angle jump*

The analysis of this sub-section implies that even though the DC-component resulted during transient conditions in the wind-farm system can be damped by the losses of the system, the SSR still remains undamped. The damping service to SSR can be provided through shunt compensation, which will be the primary focus in the following sub-section. Besides, the DC-component can also be damped/blocked by the series compensation, which will be discussed in the subsequent section. Nevertheless, the frequency and time-domain characteristics discussed here can serve as a basis for designing the compensating units.

## 2 SHUNT COMPENSATION WITH E-STATCOM AND SYNCON

Syncons, with their inherent inertia and fault current capabilities, have been traditional solutions for grid stability, but their physical characteristics restrict their performance. In contrast, advancements in converter technologies and GFM control mechanisms, STATCOMs and E-STATCOMs can offer enhanced control tunability and immediate response to grid perturbations.

As discussed in the previous section, a GFM converter can be modeled as a voltage source in series with an inductance and resistance, where the resistance represents damping and the energy is supplied or absorbed by the converter's internal storage. The converter responds to grid disturbances almost instantaneously, typically within 50 us, thanks to open-loop control, which emulates virtual impedance. This fast response enables the converter to contribute to grid stability by mimicking the effects of inertia, inductance, and damping. Note that its available energy inherently limits the extent to which it can support or correct grid phase angle deviations. The energy required to maintain transient stability depends on the collective behavior of other converters in the network that help restore power balance.

## 2.1 Integration of E-STATCOM and Syncon

The expression in (3) serves as the foundation for developing linearized models of both the E-STATCOM and the Syncon. Unlike conventional dq-reference frame modeling, this approach uses the admittance matrix in the power domain (PQ-domain). As a result, the E-STATCOM or Syncon can be easily cascaded with the grid-connected wind farm system shown in Figure 2. This enables the derivation of the overall system transfer function by simply cascading each unit for small-signal analysis, without requiring reference frame transformations.

It is essential to clarify how the E-STATCOM and Syncon control loops are modeled. The GFM-based E-STATCOM employs a swing-equation-based power control loop. Its inner control loop is assumed to be ideal and significantly faster. This assumption holds because the analysis focuses on the synchronous and sub-synchronous frequency ranges, while the inner loops primarily affect higher frequencies (typically above 100 Hz). The control also includes an open-loop virtual admittance control (VAC) with parameters $R_v = X_v = 0.2$ p.u. The power control loop has an inertia constant of H = 0.15 s and a damping factor of 0.7. Further details on the control structure are available in our previous work [11]. Note that, for a fair comparison, the E-STATCOM is designed to emulate the characteristics of a standard Syncon. Specifically, the VAC parameters are selected to match the Syncon's winding reactance and resistance, and the inertia constant and damping factor are aligned accordingly. Additionally, the Syncon's automatic voltage regulator (AVR) is modeled to operate relatively slowly..

## 2.2 Comparative performance analysis

Figure 5(a) presents a comparative analysis of the active power response of a grid-connected wind farm system under three conditions: without compensation, with shunt compensation via an E-STATCOM, and with a Syncon. As previously discussed, both the GFM E-STATCOM and the Syncon are configured with equal inertia constants to ensure a fair comparison in their response to grid phase angle modulation. The plot illustrates the amplitude and phase of the active power response when the voltage phase angle of the wind farm is modulated at sub-synchronous frequencies. Notably, the wind farm system exhibits SSR around 10 Hz. While the Syncon provides some damping, it is insufficient to suppress the resonance entirely. In contrast, the E-STATCOM significantly damps the SSR, demonstrating its effectiveness in enhancing grid stability. This behavior is further validated in Figure 5(b), which shows the time-domain response of the system to a $10^0$-phase jump. The E-STATCOM clearly mitigates oscillations more effectively than the Syncon.

What's more, the Syncon tends to attenuate the DC component of the power response rather than providing damping. The GFM E-STATCOM does not damp this DC component either, but it avoids further attenuation. The damping of this DC component can be improved by either increasing system losses (demonstrated in Figure 3), i.e., enhancing the resistive part of the impedance, or by incorporating a series compensation unit into the system.

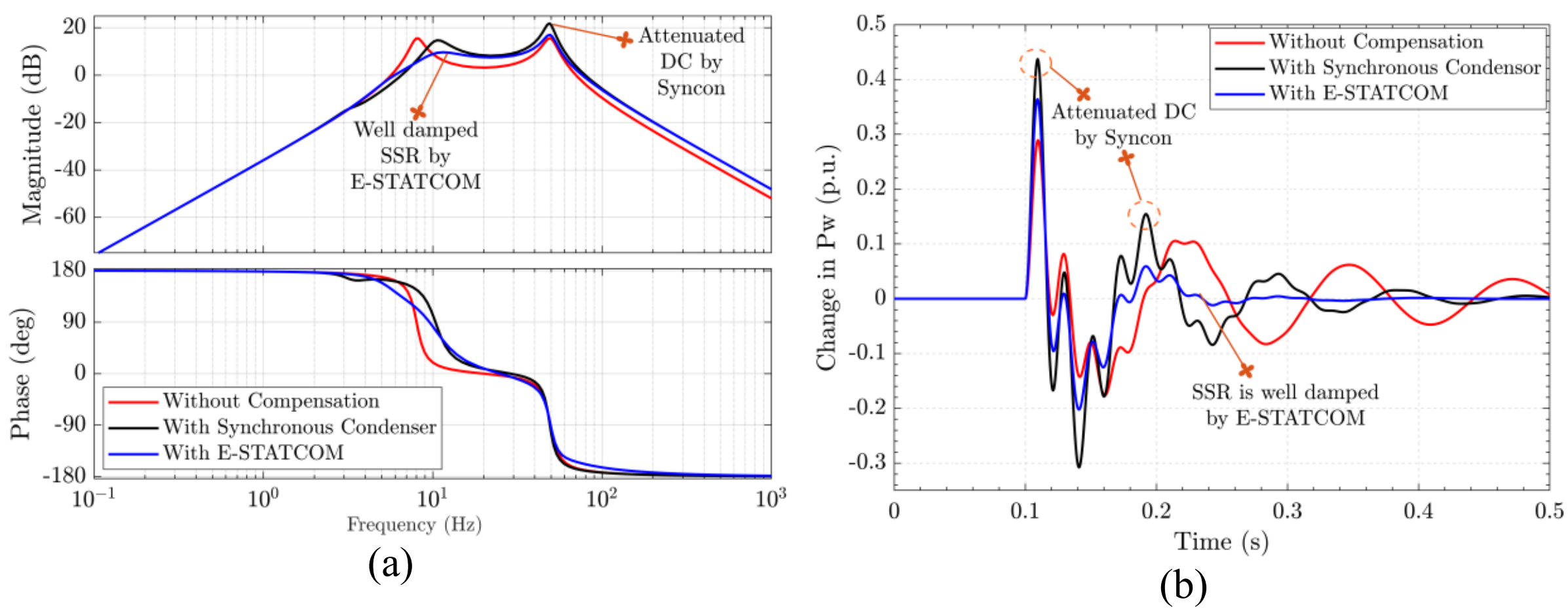

*Figure 5: Power response of the grid-connected wind farm system with shunt compensation, (a) Frequency-domain results of $P_w$ with phase perturbations, (b) Time-domain response of $P_w$ with $10^0$ angle jump*

## 3 SERIES COMPENSATION AND ITS IMPACT

Renewable energy-dominated grids often employ series compensation to enhance power transfer capability and improve system stability. Renewable sources like wind and solar are typically connected through long transmission lines, introducing significant inductive reactance. Series compensation helps reduce this reactance, thereby increasing the effective power transfer and improving voltage profiles. Moreover, it plays a critical role in damping and mitigating the fundamental frequency oscillations (DC-component) that can arise in a weak grid. However, series compensation may introduce new challenges, such as resonance at critical frequencies.

The active power response of the system with only series compensation and no shunt compensation (shown in red in Figure 6) reveals that while the DC component is blocked, new oscillatory modes emerge around 30 Hz, and a mirror resonance appears near 80 Hz. Additionally, the SSR around 10 Hz, caused by the wind farm, remains undamped.

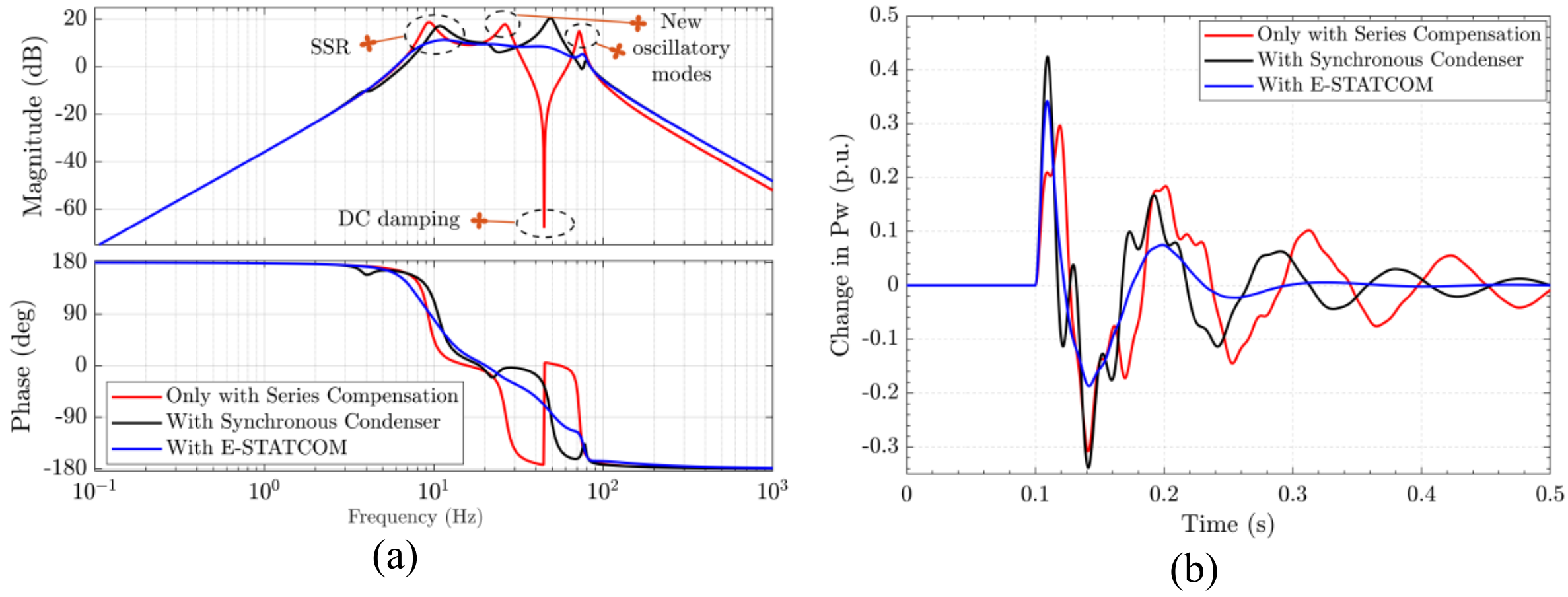

*Figure 6: Power response of the grid-connected wind farm system with both series and shunt compensation, (a) Frequency-domain results of $P_w$ with phase perturbations, (b) Time-domain response of $P_w$ with $10^0$ angle jump*

When a Syncon is added as a shunt compensation unit, it provides partial damping to the SSR but reintroduces the DC component that was previously suppressed by the series compensation.

Although the new oscillatory modes are somewhat damped, they still exhibit noticeable peaks. In contrast, the E-STATCOM, when used as a shunt compensator, offers damping across the entire frequency range. This is clearly illustrated by the flat response curve in blue in Figure 6(a). These observations are further confirmed through time-domain simulations. As shown in Figure 6(b), the oscillations in the wind power due to a $10^0$ angle jump (shown in red color response) are well damped by the E-STATCOM (shown in blue).

Note that the above linear modeling analyses and results assume an ideal, unsaturated Syncon model. Accounting for winding/reactance and transformer saturation may further degrade Syncon's performance.

## 4 SIL WITH DETAILED PSCAD MODEL

Simulation-in-loop (SIL) is conducted within the EMT environment to validate the theoretical concepts and mathematical models previously discussed. The system under study incorporates a detailed network featuring realistic transformer, converter, synchronous machine, and wind farm models, as illustrated in Figure 1. The generating source (wind farm) and shunt compensating units are connected to the 275 kV PCC bus, while the grid voltage is maintained at 400 kV. It should be noted that the wind farm connects to the PCC bus via long-distance cables, which may result in SSR occurring at a frequency different from that identified in linear modeling. Nonetheless, the nature and effect of SSR remain consistent, necessitating effective damping measures.
Compensating devices, such as Syncons or GFM E-STATCOMs, are essential to address various operational challenges and stabilize the grid in this configuration. The following analysis evaluates their performance across different transient scenarios. To ensure comparability, both devices are set with an inertia constant of 2 s and a sub-transient reactance of 20 %. All time-domain results presented in the subsequent sub-sections are expressed in p.u. values based on the PCC bus parameters.

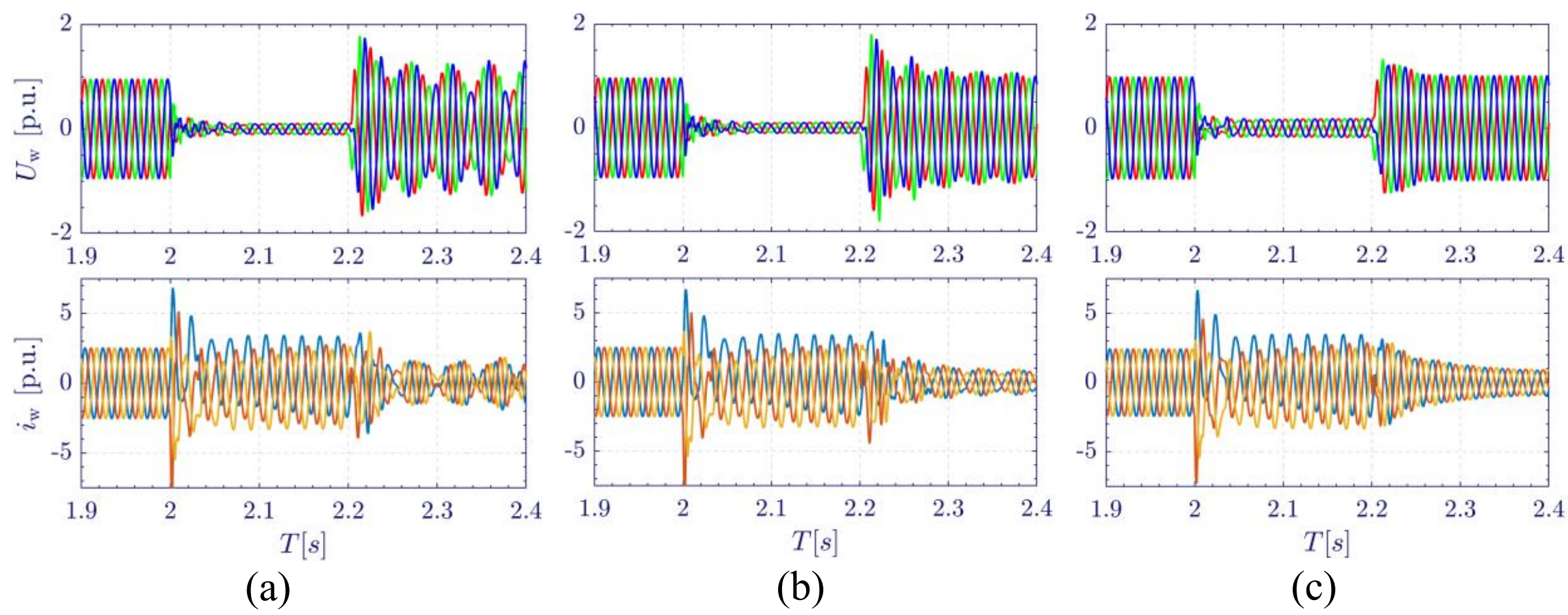

*Figure 7: Voltage and current waveforms of the wind farm under three-phase to ground fault (a) without shunt compensation, (b) Compensation with Syncon, (c) Compensation with E-STATCOM*

### 4.1 Analysis of SSR

The grid-connected wind-farm is subjected to a three-phase-to-ground fault scenario, during which the compensating devices are temporarily disconnected to produce the results in Figure 7(a). The fault initiates at 2 s and is cleared after 0.2 s. Following fault clearance, the wind bus voltage ($U_w$) exhibits a pronounced transient overvoltage (TOV) reaching 2 p.u., along with an SSR at 23 Hz. Figure 7(b) shows that Syncon offers some damping to the SSR, which is not fast enough. In comparison, the E-STATCOM demonstrates superior performance by effectively suppressing the TOV, suppressing the DC component, and damping the SSR within a single cycle.

## 4.2 Analysis of DC component

In power systems with a high share of renewable energy, phase angle shifts—typically between 10° and 30°—occur frequently. These shifts result in DC components that require damping to ensure system stability. Figure 8 demonstrates the system's response to a +30° phase jump. In Figure 8(a), no shunt compensation unit is connected, so its power contribution is zero. Figure 8(b) presents the Syncon's response, indicating that the Syncon injects a DC component, which is more prominent than the SSR, and displays a slower damping response as observed from the injected power. In comparison, the E-STATCOM shows a faster reaction and suppresses the DC component more effectively in Figure 8(c).

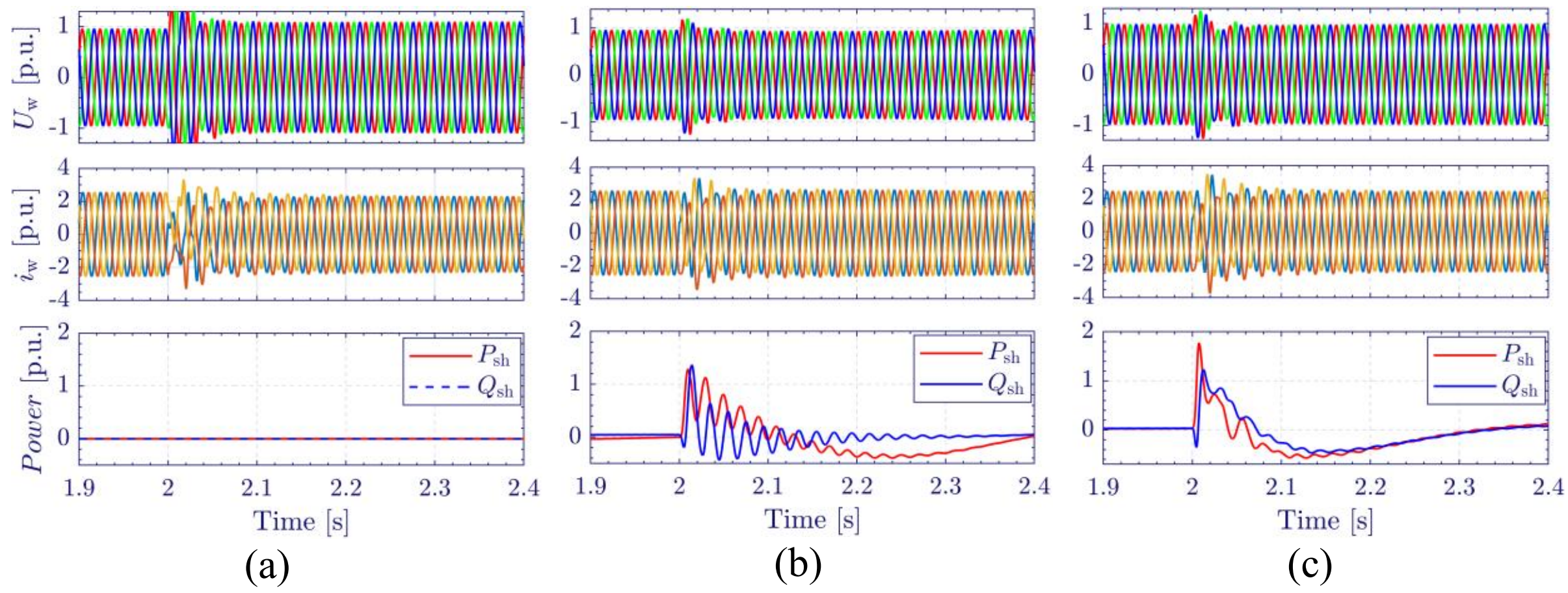

*Figure 8: Voltage and current waveforms of the wind farm and power from the compensating unit under $30^0$ phase jump (a) without shunt compensation, (b) with Syncon, (c) with E-STATCOM*

## 4.3 Analysis of overcurrent capability

A single-phase-to-ground fault is introduced at 2 s and lasts for 0.2 s, as depicted in Figure 9. It can be observed that the Syncon injects fault current without restriction, as its response is governed by sub-transient reactance. On the other hand, the E-STATCOM's fault current output is constrained by its valve capacity, as presented in Figure 9(b). However, Figure 9(c) demonstrates that an E-STATCOM designed with overcurrent handling capabilities can deliver fault current levels comparable to those of Syncon. Notably, during the post-fault recovery period, the E-STATCOM is capable of maintaining full current support—something Syncons are unable to do. This overcurrent capability could be achieved through special design considerations in the converter, such as temporary overload of the converter or through specially designed semiconductors incorporating very short-term overcurrent capability.

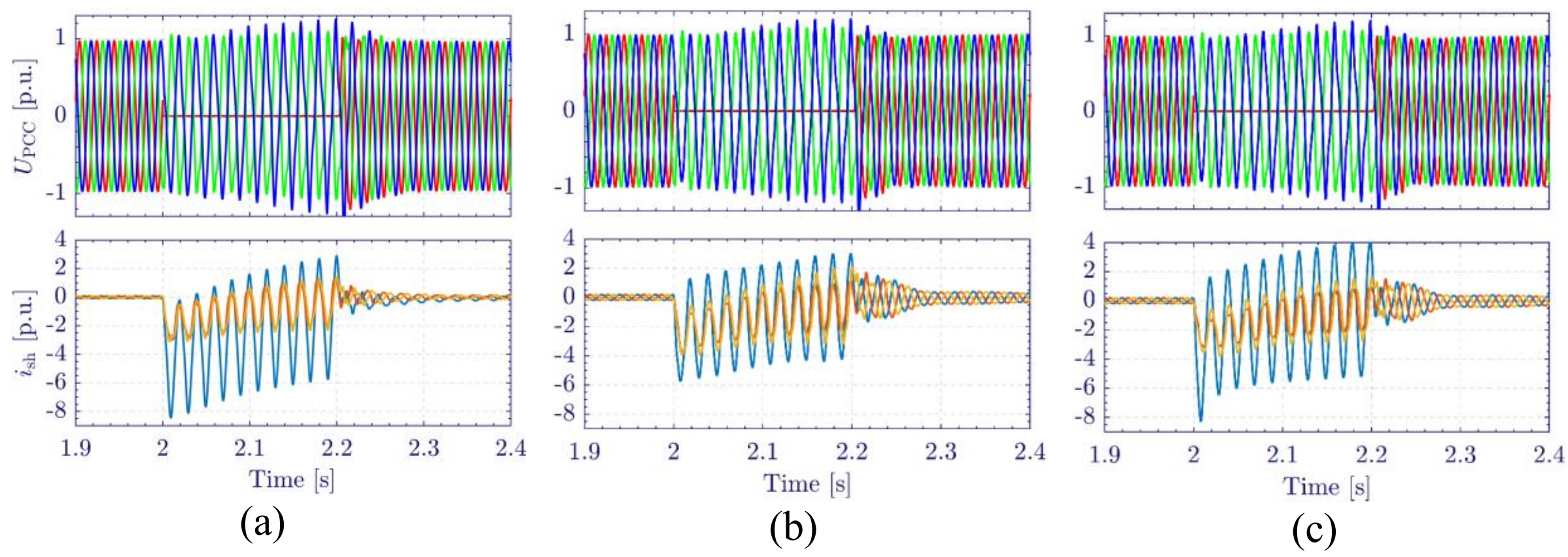

*Figure 9: Voltage at PCC and fault current contribution under line-ground fault (a) with Syncon, (b) with E-STATCOM, (c) with E-STATCOM having overcurrent capability*

## 5 CONCLUSION

This paper first introduced a power-admittance-based linear modeling framework for grid-forming E-STATCOMs, designed to be intuitive, reference-frame-independent, and scalable for integrating converter-based resources. This analytical foundation enables clear characterization of GFM-based E-STATCOM behavior in different dynamic grid conditions and facilitates comparison with synchronous condensers across both frequency and time domains.

Building on this framework, the second half of the paper presented detailed EMT simulations to validate the theoretical insights. These simulations demonstrated the superior damping capability of E-STATCOMs in mitigating sub-synchronous resonance and DC components, as well as their effectiveness in suppressing transient overvoltages. In addition, the E-STATCOM effectively provides damping service to the new oscillatory modes that emerge due to series compensation, ensuring stable operation across a wide frequency range. Furthermore, the E-STATCOM's overcurrent capability was shown to match the fault current contribution of synchronous condensers while offering enhanced support during the recovery phase. Overall, the study confirms that E-STATCOMs offer a more effective and targeted solution than synchronous condensers for RES systems with long cables and series compensation, and outlines both the rationale and design approach needed to achieve robust grid performance.